\documentclass[jcp,twocolumn,superscriptaddress,floatfix,graphicx,showpacs]{revtex4-1}
\usepackage{ifpdf}
 \newif\ifpdf
\ifx\pdfoutput\undefined
   \pdffalse
\else
   \pdfoutput=1
   \pdftrue
\fi
\ifpdf
   \usepackage{graphicx}
   \usepackage{epstopdf}
   \usepackage{color}
\else
   \usepackage{graphicx}
\fi
\usepackage{cancel}
\usepackage{srctex}%\usepackage{hyperref}
\usepackage{hyperref}

\usepackage{amsmath,amssymb}

\usepackage{epsfig}

\usepackage{epsfig}

\begin{document}

\title{Simulated Cu$\--$Zr glassy alloys: the impact of composition on icosahedral order}

\author{B.A. Klumov}
\affiliation{Aix-Marseille-Universit\'{e}, CNRS, 13397 Marseille, France}
\affiliation{High Temperature Institute, Russian Academy of Sciences, 125412, Moscow, Russia}
\affiliation{L.D. Landau Institute for Theoretical Physics, Russian Academy of Sciences, 117940 Moscow, Russia}

\author{R.E. Ryltsev}
\affiliation{Institute of Metallurgy, Yekaterinburg 620016, Russia}
\affiliation{Ural Federal University, Yekaterinburg 620002, Russia}
\affiliation{L.D. Landau Institute for Theoretical Physics, Russian Academy of Sciences, 117940 Moscow, Russia}

\author{N.M. Chtchelkatchev}
\affiliation{L.D. Landau Institute for Theoretical Physics, Russian Academy of Sciences, 117940 Moscow, Russia}
\affiliation{Moscow Institute of Physics and Technology, 141700 Moscow, Russia}
\affiliation{ All-Russia Research Institute of Automatics, Moscow 127055, Russia}
\affiliation{Institute of Metallurgy, Yekaterinburg 620016, Russia}

%\date{\today}

\begin{abstract}
The structural properties of the simulated $\rm Cu_{\alpha}Zr_{1-\alpha}$ glassy alloys
are studied in the wide range of the copper concentration $\alpha$ to clarify the impact of the composition on the number density of the
icosahedral clusters. Both bond orientational order parameters  and Voronoi tessellation methods are used to identify these clusters. Our analysis shows that abundance of the icosahedral clusters and the chemical composition of these clusters are essentially non monotonic versus $\alpha$ and demonstrate local extrema. That qualitatively explains the existence of pinpoint compositions of high glass-forming ability observing in Cu$\--$Zr alloys. Finally, it has been shown that Voronoi method overestimates drastically the abundance of the icosahedral clusters in comparison with the bond orientational order parameters one.
\end{abstract}


\maketitle

\section{Introduction}

Cu$\--$Zr system is in short list of binary alloys capable to form bulk-metallic glasses (BMG)\cite{Xu2004ActMat,Wang2004AppPhysLett,Wang2005JMatRes}. It has unique glass-forming ability (GFA) where compositions of BMG formation are located in narrow concentration intervals (so-called pinpoint compositions)~\cite{Wang2004AppPhysLett,Li2008Science,Yang2012PRL}.  One of the general ideas explaining this unusual property, strongly supported by molecular dynamics simulations, is the suggestion that local
icosahedral order plays important role in GFA \cite{Li2009PRB,Peng2010ApplPhysLett,Soklaski2013PRB,Wu2013PRB,Wen2013JNonCrystSol,Meng2015PhysChemLiq,Wang2015JPhysChemA}. However the definition of local order contains a significant proportion of uncertainty related to the complexity of the definition of the nearest neighbors in disordered media (see e.g. \cite{SANN,Malins2013}) and the difficulty of symmetry-classification of local clusters with non-ideal shape. This problem makes the results of local order analysis rather challenging for practical applications.

Here we study structural properties of the simulated $\rm Cu_{\alpha}Zr_{1-\alpha}$ glassy alloys in the wide range of the copper concentration $\alpha\in(0.2,1)$ using two different methods to detect the local order: Voronoi tessellation (VT) \cite{VT} and Bond Orientational Order Parameters (BOOP) \cite{Steinhardt1981}. We show that more rigorous BOOP method reveals essentially different results in comparison to those obtained by widely accepted VT method. Comparison of the BOOP and VT demonstrates that the latter significantly overestimates the number of icosahedral-like (ico-like) particles in the system. We observe that both the abundance of
ico-like clusters and their chemical composition vary non-monotonically with concentration. Besides, chemical composition of
ico-like clusters is found to be deviated significantly from the mixing ratio at $\alpha > 0.25 $; relative lacks of copper  atoms vary in the range $6 \div 12 \%$ with non monotonic dependence on $\alpha$.

\begin{figure}
  \centering
  % Requires \usepackage{graphicx}
  \includegraphics[width=0.45\textwidth]{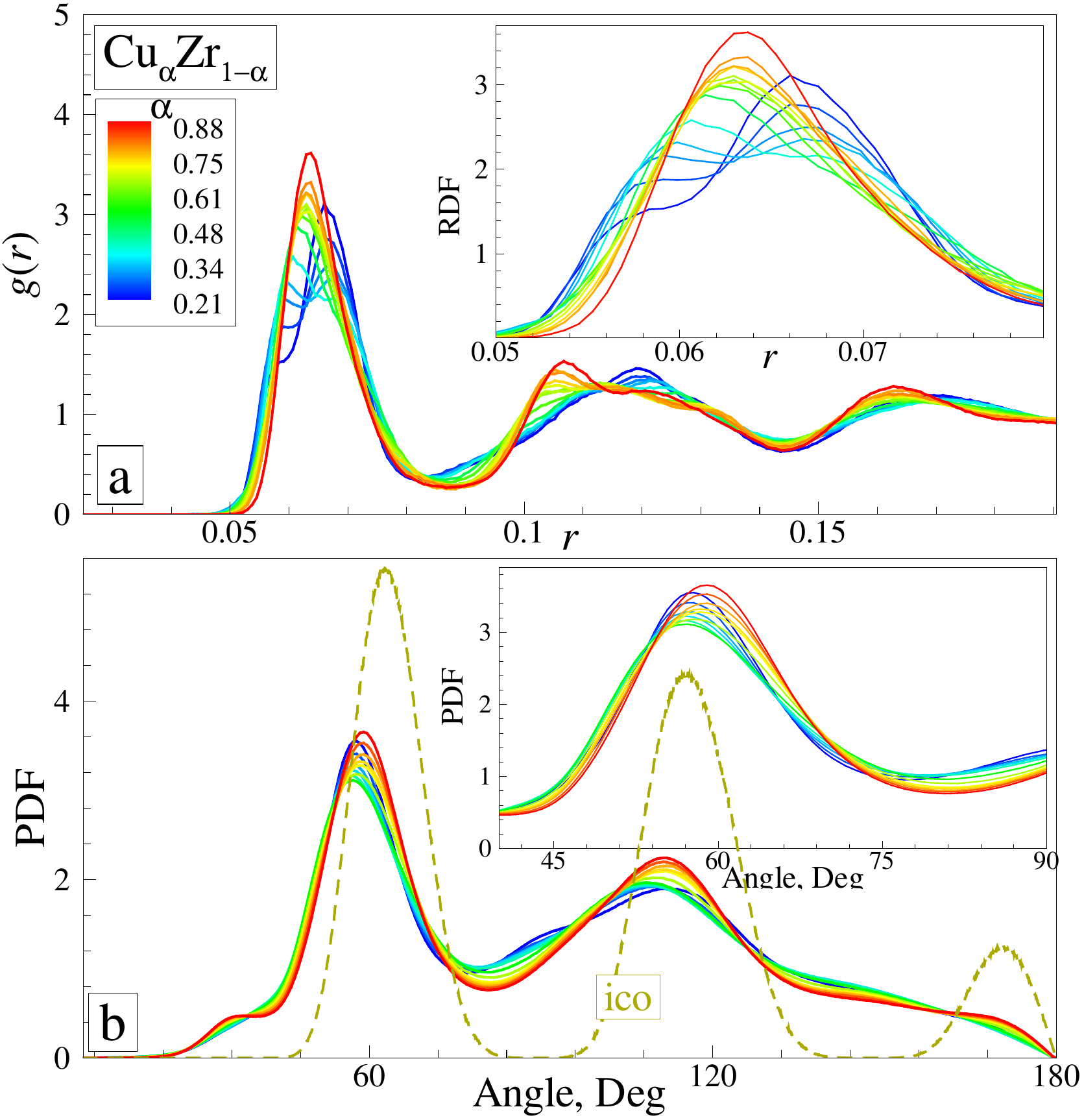}
  \caption{(Color online). Two-point correlations in glassy $\rm Cu_{\alpha}Zr_{1-\alpha}$ system. The total radial distribution function $g(r)$ (a) and the bond angle distribution function
  (BADF) (b) (calculated by using 12 nearest neighbors) are shown at different copper abundance $\alpha$. Insets show the fine details of the first peak of these distributions.
  Additionally, BADF for the icosahedral cluster (with thermal vibrations taking into account)
  is plotted by dashed line. Temperature of the system is $T=300$ K. The cooling rate $\gamma$ is $10^{11}$ K/s.}
  \label{fig1}
\end{figure}

\section{Methods}

For the molecular dynamics (MD) simulations, we used LAMMPS package \cite{Plimpton1995JCompPhys}. The system of $N \approx 5000$ particles was simulated under periodic boundary conditions in Nose-Hoover NPT ensemble at pressure $P=1$~atm. The MD time step was 1~fs. Initial configurations were equilibrated melts
at $T = 1200$~K. The system was cooled down to $T=300$~K with the cooling rate $\gamma = 10^{11}$~K/s. This rate was chosen to compare our results with those obtained by other authors who had used this $\gamma$ value (see inset in Fig.~\ref{fig4}). Though in general the structure of the glass can be essentially cooling rate dependent \cite{Vollmayr1996JCP,Zhang2014AppPhysLett,Yanjun2015TrNonFerMetChin,Ryltsev2016}, it is not important for this study where composition dependence of the structural properties is investigated. Interactions between alloy components is described here using widely accepted embedded atom potential specially adopted for liquid and glassy Cu$\--$Zr system~\cite{Mendelev2009PhilMag}.

The BOOP method~\cite{Steinhardt1983} has been widely used in condensed matter physics to quantify the local orientational order~\cite{Steinhardt1981,Steinhardt1983,Mitus1982,Trus2000} in Lennard-Jones and hard sphere systems~\cite{ErringtonLJ,RT96,Luchnikov2002,Jin2010,KlumovLJ13a,KlumovLJ13b,Bar14,KLumovJPC14},
complex plasmas~\cite{CPP04,RZ06,JETP08,JETPL09,PPCF09,KlumovPU10,Khrapak2012},
colloidal~\cite{Gasser01,Kawasaki10} and patchy systems~\cite{Vasilyev2013,Vasilyev2015}, etc.
The method allows us to explicitly recognize symmetry of local atomic clusters~\cite{KlumovPU10,KlumovPRB11,Hirata2013Science} and study their spatial distribution~\cite{Ryltsev2013PRE,Ryltsev2015SoftMatt}.

\begin{figure}
\centering
% Requires \usepackage{graphicx}
\includegraphics[width=0.45\textwidth]{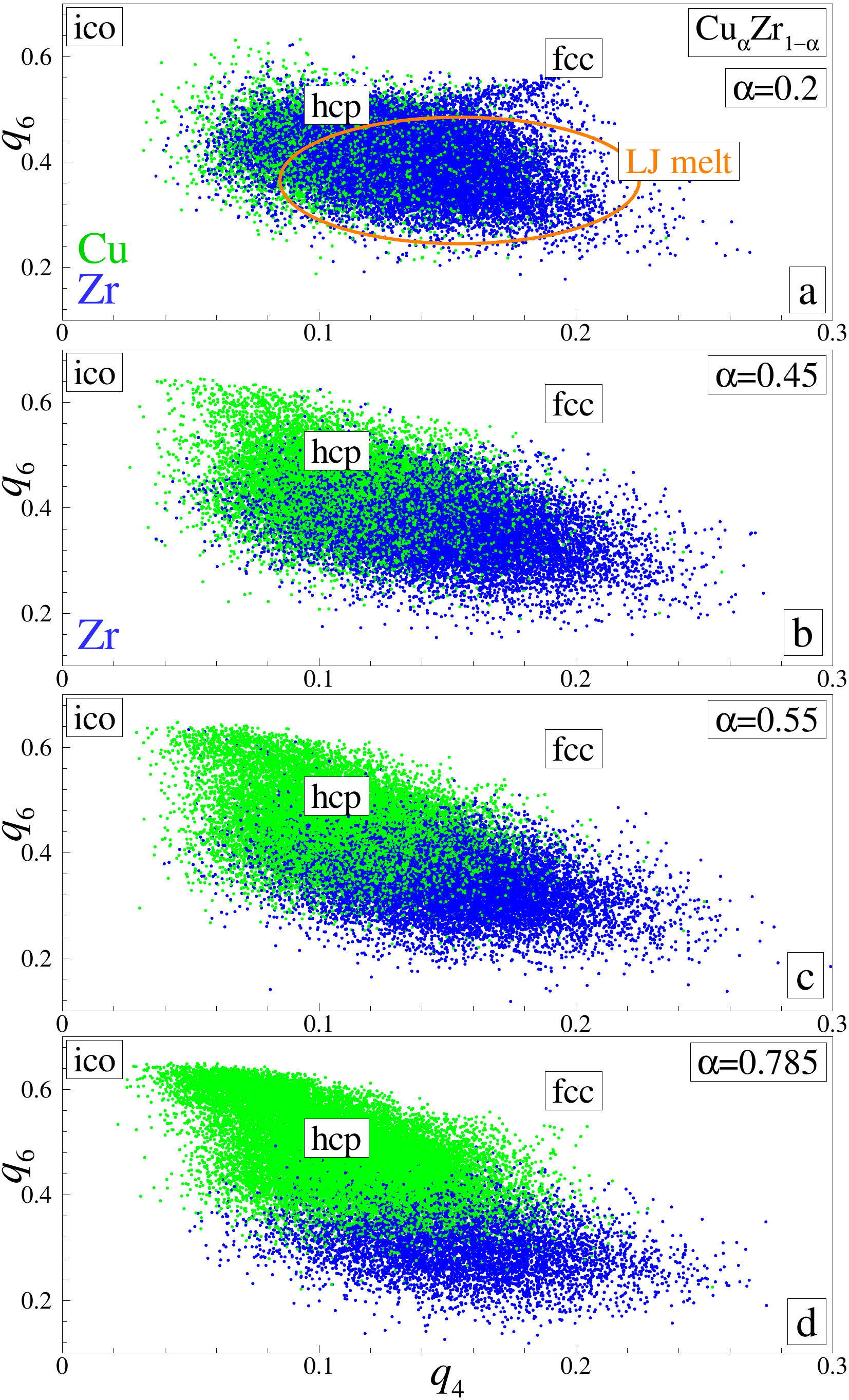}\\
\caption{(Color online). Local orientational order of the ${\rm Cu_{\alpha}Zr_{1-\alpha}}$ system on the plane $q_4\--q_6$ at different Cu abundance $\alpha$. Bond orientational order parameters (BOOP) were calculated via 12 nearest neighbors for both copper (green) and zirconium (blue) atoms to identify different close packed structures. The formation of Cu-centered ico-like clusters is clearly seen at high values of $\alpha$; zirconium atoms are nearly completely in disordered liquid-like phase.
BOOP for the perfect icosahedron, hcp and fcc clusters are also indicated for the comparison, as well as the Lennard-Jones melt distribution bounds (panel (a), orange line). Temperature of the system is $T=300$ K.}
\label{fig2}
\end{figure}

\begin{figure}
\centering
% Requires \usepackage{graphicx}
\includegraphics[width=0.45\textwidth]{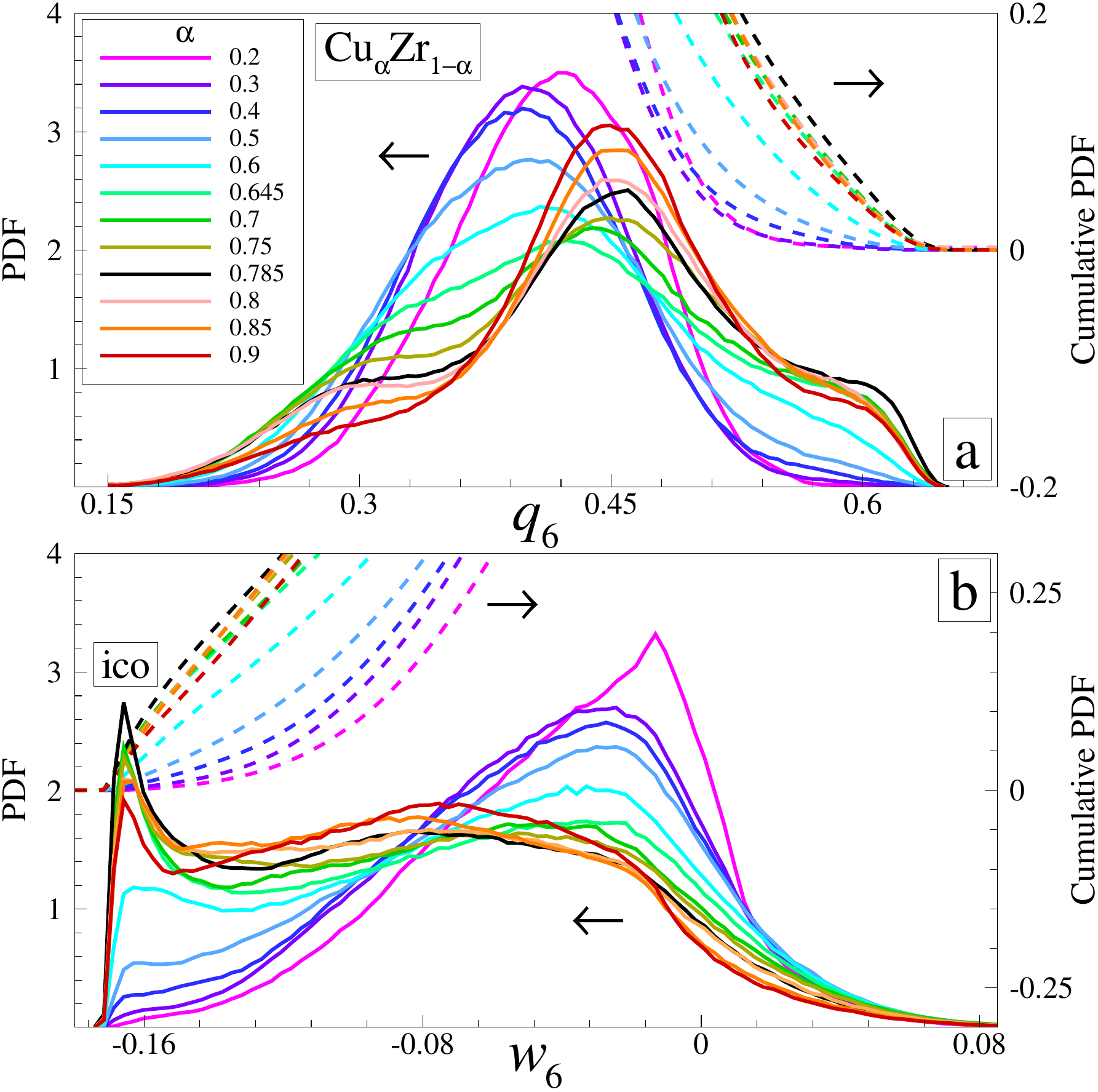}\\
\caption{(Color online). Local orientational order of the ${\rm Cu_{\alpha}Zr_{1-\alpha}}$ system taken at $T=300$ K: the probability distribution functions $P$ over rotational invariants $q_6$ (a) and $w_6$ (b) at different copper abundances $\alpha$. Cumulative distributions ($C_q(x)=\int_{\infty}^x P(q_6)dq_6$ and $C_w(x)=\int_{-\infty}^x P(w_6)dw_6$) at different $\alpha$ are also shown to evaluate the density of atoms with
ico-like symmetry (having $q_6 > 0.6$ and $w_6 < 0.15$). Increase of $\alpha$ value up to moderate values ($\alpha
\simeq 0.5$) clearly results in increase of ico-like clusters in the system; the increase is not monotonic then (with the maximum of the ico-like atoms at $\alpha = 0.785$ (black line)).}
\label{fig3}
\end{figure}

\begin{figure}
\centering
% Requires \usepackage{graphicx}
\includegraphics[width=0.45\textwidth]{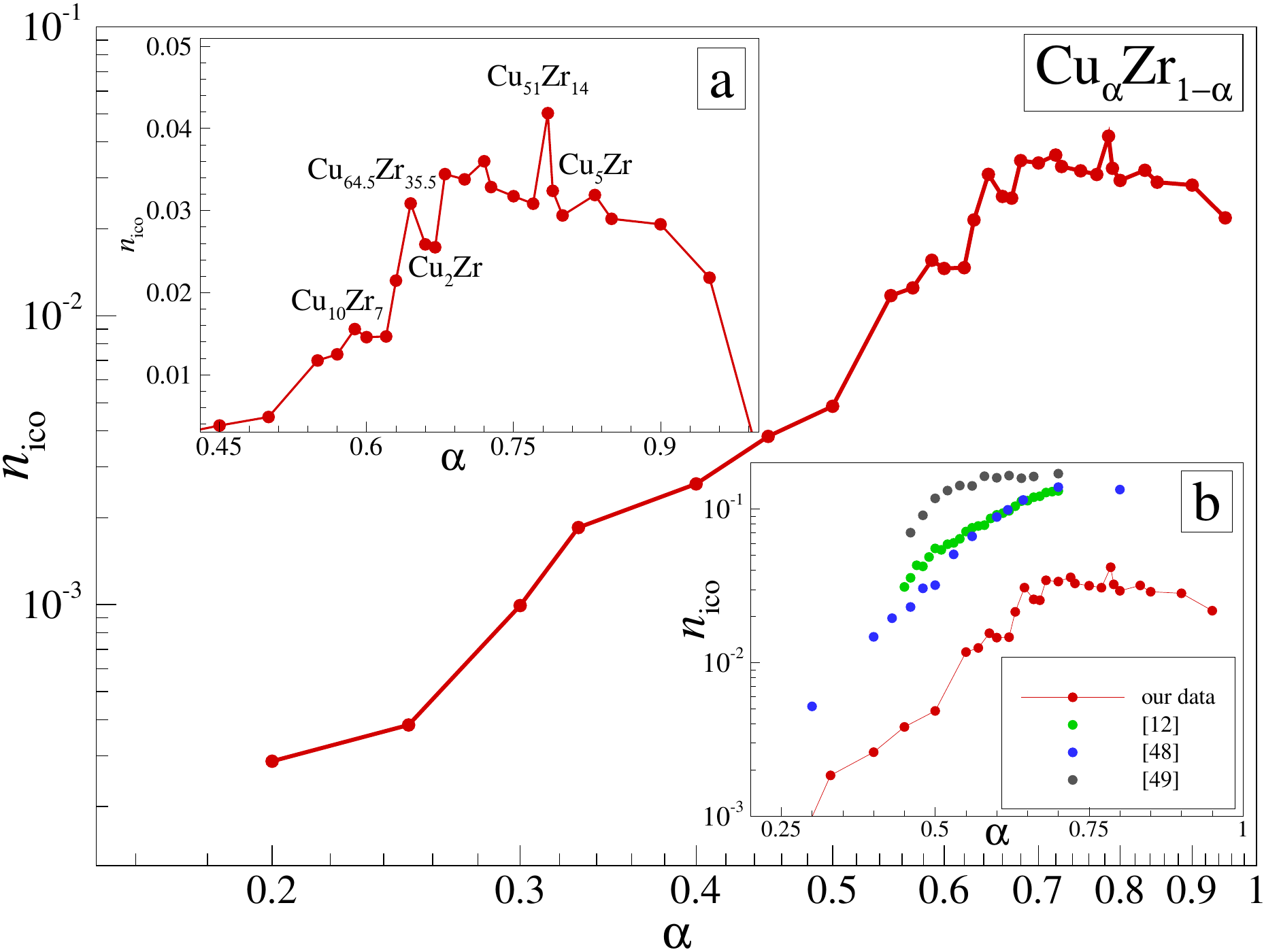}\\
\caption{(Color online). Simulated ${\rm Cu_{\alpha}Zr_{1-\alpha}}$ alloy at $T=300$ K.
Main frame: abundance of ico-like clusters $n_{\rm ico}$ (calculated via cumulative functions $C_q(q_6)$ and $C_w(w_6)$ presented in Fig.~\ref{fig3}) versus copper abundance $\alpha$. Non monotonic behavior of the ico-like cluster density is seen at $\alpha \ge 0.5$. Left inset (a) shows the fine details of $n_{\rm ico}(\alpha)$ dependence in the range $\alpha>0.45$.  Right inset (b) shows comparison of our $n_{\rm ico}(\alpha)$ dependence with those obtained by other authors \cite{Ward2013PRB,Lad2014JNonCrystSol,Wang2015JPhysChemA}.}
\label{fig4}
\end{figure}

For BOOP investigation we define the rotational invariants (RI) of rank $l$ of the second order $q_l$ and third order $w_l$~\cite{Steinhardt1981,Steinhardt1983}. The advantage of $q_l$ and $w_l$ is that they are uniquely determined for any polyhedron including the elements of any crystalline structure. Among the RI, $q_4$, $q_6$, $w_4$, $w_6$ are typically the most informative ones so we use they in this work. To identify
ico-like clusters we calculate the rotational invariants $q_l$ and $w_l$ for each atom using the fixed number of nearest neighbors ($N_{\rm nn}=12$). Atom whose coordinates in the space $(q_4, q_6, w_4, w_6)$ are sufficiently close to those of the perfect structures is counted as ico-like (fcc-like, hcp-like) etc. The RI for a number of crystalline structures are shown in Table~\ref{table1}.

\begin{table}[th]
\centering
\caption{Rotational invariants $q_l$ and $w_l$ ($l=4,~6$) of a few perfect crystalline clusters calculated via 12 nearest neighbors:
icosahedron (ico), face centered cubic (fcc) and hexagonal close-packed (hcp).}
\begin{tabular}{|c|c|c|c|c|}
%\begin{tabular}{ll|cccc}
\hline %\hline
cluster type & \quad $q_{4}$ & \quad $q_{6}$ & \quad $w_{4}$ & \quad $ w_{6}$
\\ \hline ico  & $1.4 \times 10^{-4}$ & 0.6633 & -0.1593  & -0.1697
\\ \hline fcc  & 0.1909  & 0.5745  & -0.1593 &  -0.0131
\\ \hline hcp  & 0.0972 & 0.4847 & 0.1340  & -0.0124
\\\hline %\hline
\end{tabular}
\label{table1}
\end{table}

\section{Results}

Fig.~\ref{fig1} shows two-point correlations in glassy $\rm Cu_{\alpha}Zr_{1-\alpha}$ alloys at
$T=300$~K. The total radial distribution function $g(r)$ (a) and the bond angle distribution function (BADF)(b) are shown at different copper abundance $\alpha$ in the range $0.2 \div 0.95$. The shoulder on the first peaks of $g(r)$ revealing the chemical-like order is clearly observed only at small $\alpha$; splitting of the second peak (which is widely used as the indicator of the glassy state and the appearance of the
ico-like order in the system) occurs at moderate $\alpha > 0.5$. BADFs shapes at different $\alpha$ indicate possible presence of ico-like clusters.

Fig.~\ref{fig2} shows the local orientational order of the glassy ${\rm Cu_{\alpha}Zr_{1-\alpha}}$
alloys on the plane of rotational invariants $q_4\--q_6$ at few $\alpha$ values. Atoms distribution on the plane clearly shows the formation of Cu-centered
ico-like clusters at moderate and big $\alpha$ (panels b), c), d)) and mostly disordered-like behavior (with traces of hcp-like and fcc-like clusters) at small $\alpha$ (panel a).

To quantify the distributions in Fig.~\ref{fig2}, we use the normalized
one-dimensional probability density function over different rotational invariants $P(q_l)$ and $P(w_l)$, so that (for e.g. $q_l$) we have $\int_{-\infty}^{\infty} P(q_l)dq_l \equiv 1$.
Even more convenient is using (cumulative) distribution functions associated with the $P(q_l)$ and $P(w_l)$~\cite{KlumovPU10,KlumovPRB11} which are defined as (by using as example the bond order parameter $q_6$): $C_q(x)=\int_{-\infty}^x P(q_6)dq_6$. Using the cumulative functions we can find abundance (density) of any solid-like structure with given accuracy $\delta_{\rm cr}$: $C_q(q^{\rm cr}+\delta_{\rm cr})$ - $C_q(q^{\rm cr}-\delta_{\rm cr})$, where $q^{\rm cr}$ is the bond order parameter for the given lattice type.

We note, that the set of distributions $P$ and $C$ taken for different $q_l$ and $w_l$ completely describes the local orientational order in the system as an abundance of different ordered and disordered structures. Fig.~\ref{fig3} shows how the distributions $P(q_6)$ and $P(w_6)$ and the corresponding cumulative functions $C_q$ and $C_w$ vary with the increase of $\alpha$. Behaviour of the distributions  $P(q_6),~P(w_6)$ and $C(q_6),~C(w_6)$ in the vicinity of the perfect
icosahedron (having $q_6^{\rm ico} \approx 0.66$ and $w_6^{\rm ico} \approx -0.17$ (see, Table~\ref{table1})) clearly shows non monotonic dependence of density of ico-like atoms on $\alpha$ value. For instance, at considered cooling rate ($\gamma = 10^{11}$ K/s), the maximum density of ico-like clusters occurs at $\alpha = 0.785$. To better illustrate such non monotonic behaviour, we show in Fig.~\ref{fig4} the dependence of the abundance of ico-like clusters $n_{\rm ico}$ on Cu concentration $\alpha$. A cluster is treated as ico-like one if it has order parameters $q_6 > 0.6$ and $w_6 < -0.16$. We see that $n_{\rm ico}(\alpha)$ is essentially non monotonic with maximal amount of ico-like clusters within the range $0.6<\alpha< 0.9$. Moreover, $n_{\rm ico}(\alpha)$ demonstrates local extrema located at stoichiometry compositions of a few Cu-Zr intermetallic compounds as well at the BMG composition ${\rm Cu_{0.645}Zr_{ 0.355}}$. The presence of such extrema correlates with the existence of pinpoint compositions of BMG formation in real Cu$\--$Zr alloys. Indeed, the ranges of local $n_{\rm ico}(\alpha)$ maxima are expected to correspond to high GFA compositions. The very fact that simple EAM model potential can describe such non-trivial behaviour (at least qualitatively) is the important result. Moreover, the maximum at  ${\rm Cu_{0.645}Zr_{ 0.355}}$ composition exactly corresponds to one of the pinpoint composition of BMG formation. So even the quantitative agreement with experimental data takes place.

\begin{figure}
\centering
% Requires \usepackage{graphicx}
\includegraphics[width=0.45\textwidth]{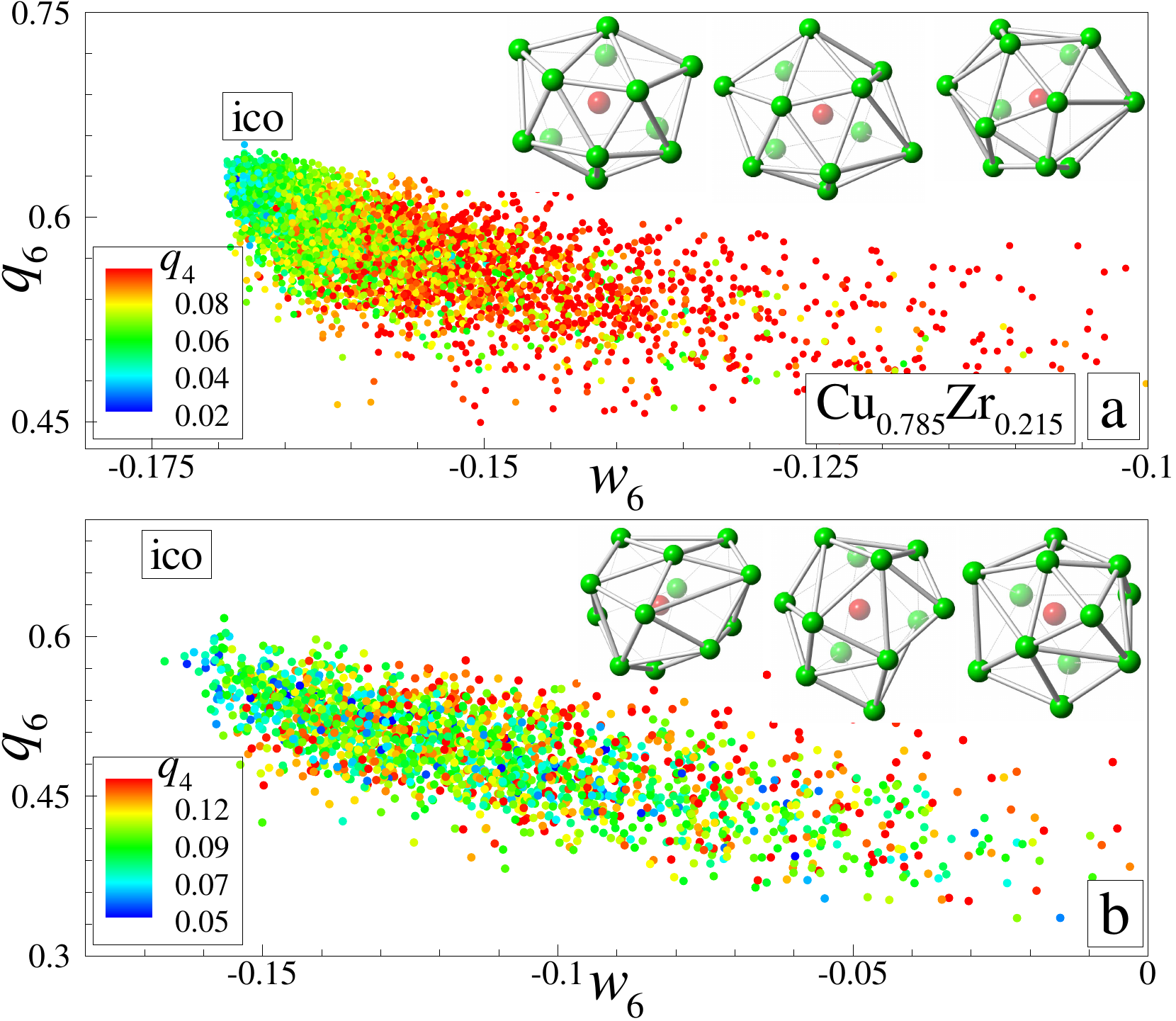}\\
\caption{(Color online). Alloy ${\rm Cu_{0.785}Zr_{0.215}}$ at $T=300$ K. Comparison of the
Voronoi and BOOP method for the ico-like cluster identification. Distribution of the particles counted as full
icosahedra (i.e. having Voronoi index $\langle 0,0,12,0\rangle $) (a) and distorted icosahedra (having Voronoi index $\langle 0,2,8,2\rangle$) (b) on the plane or rotational invariants $w_6 \-- q_6$. Particles are color-coded via $q_4$ value. Insets show corresponding ico-like clusters (with red-colored central atom).}
\label{fig5}
\end{figure}

In the right inset of Fig.~\ref{fig4} we also show the comparison of our data with those obtained in Refs.~\onlinecite{Ward2013PRB,Lad2014JNonCrystSol,Wang2015JPhysChemA}. As seen from the picture, our $n_{\rm ico}(\alpha)$ curve demonstrates qualitatively the same behaviour as those obtained by other authors. But, quantitatively, our $n_{\rm ico}$ values are much lower then those presented in Refs.~\onlinecite{Ward2013PRB,Lad2014JNonCrystSol,Wang2015JPhysChemA}. The main reason for such deviation is
the difference of methods used to determine ico-like clusters. Our results were obtained by BOOP method but the others by VT one. The latter is the most accepted method to study short and medium range order of metallic alloys~\cite{Finney1970ProcRoySoc,Cheng2011ProgMateSci}. The main point of the method is dividing the space into polyhedra which faces are planes bisecting the lines connecting each particle with its nearest neighbors. The symmetry of polyhedra is determined by Voronoi indices like $\langle n_3,n_4,n_5,n_6\rangle$ where $n_i$ is the the number of i-edged faces.
Voronoi index is an topologically stable characteristic which is not sensitive to deformation of the clusters. For example, any polyhedron with the index  $\langle 0,0,12,0\rangle$ is treated as an icosahedron regardless of its distortion degree. That can cause the false determination of clusters structure and their abundances. That can also be the reason for the absence of local extremal in $n_{\rm ico}(\alpha)$ curves obtained by VT method.

To show that, we compare the results obtained by both methods for ico-like clusters in ${\rm Cu_{0.785}Zr_{0.215}}$ alloy. Top panel in Fig.~\ref{fig5} presents the distribution of particles counted by VT method as full
icosahedra, i.e. the particles with the Voronoi index $\langle 0,0,12,0\rangle$ on the plane of rotational invariants $w_6 \-- q_6$. Bottom panel in Fig.~\ref{fig5} shows the same distribution taken for the clusters with Voronoi index $\langle 0,2,8,2\rangle$ usually treated as ''distorted icosahedra''. Particles are color-coded via $q_4$ value. From the data presented in panel (a), we see that significant part of particles have $q_6 < 0.6$ and $w_6 > 0.16$, i.e. these particles can not be counted as ico-like particles if we use BOOP method. Although these clusters look like strongly distorted icosahedrons, some of the clusters are actually distorted hcp-like and fcc-like clusters, not ico-like ones. As to particles shown in bottom  panel (which are counted by Voronoi method as distorted icosahedrons), nearly all of them are quite far from even distorted icosahedron; these particles have $q_6$ and $w_6$ values which are typical for intermediate phase in between icosahedral and hcp/fcc type of symmetry.

\begin{figure}
\centering
% Requires \usepackage{graphicx}
\includegraphics[width=0.45\textwidth]{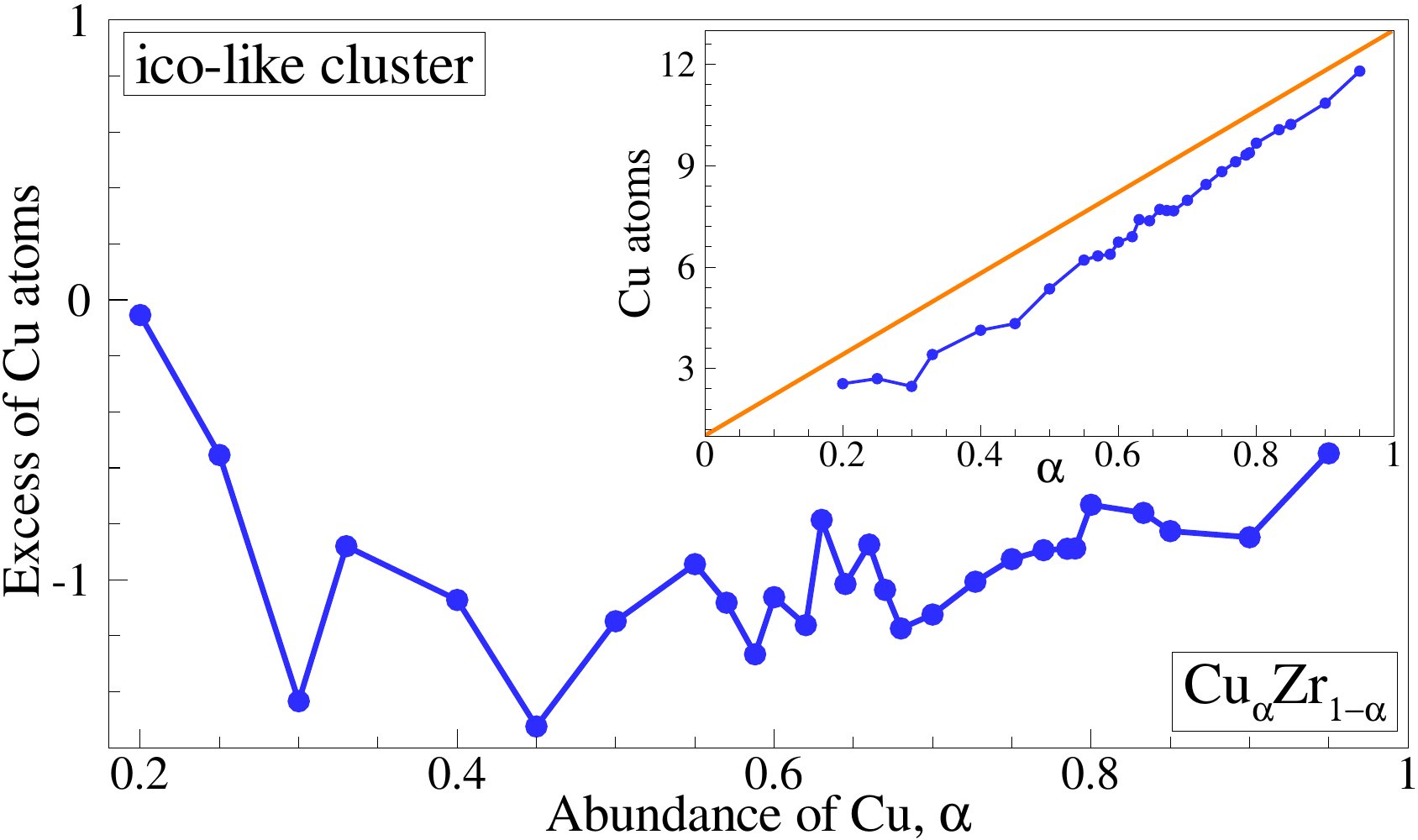}\\
\caption{(Color online). Mean chemical composition of the ico-like clusters observed
in the simulated ${\rm Cu_{\alpha}Zr_{1-\alpha}}$ system (taken at $T=300$ K).
Presented is the mean excess of Cu atoms in the cluster (with the $\alpha$ value being subtracted)
versus copper abundance $\alpha$. Inset shows how the mean number of Cu atoms in the clusters deviates from the mixing ratio $\alpha$ (orange line). Lack of Cu atoms in the ico-like cluster occurs at $\alpha > 0.2$; the lack non monotonically depends on the $\alpha$ value.}
\label{fig6}
\end{figure}

Besides the symmetry of local clusters, their chemical composition may be important to understand structure of multi-component glasses~\cite{Malins2013JCP,Wang2015JPhysChemA}. In Fig.~\ref{fig6} we present the mean chemical composition $N_{\rm Cu}(\alpha)$ of the ico-like clusters. In fact the mean excess concentration of Cu atoms in the cluster $\delta N_{\rm Cu}(\alpha) \equiv N_{\rm Cu}(\alpha) - \alpha$ is plotted versus $\alpha$. The inset shows how the value $N_{\rm Cu}(\alpha)$ deviates from the mixing ratio $\alpha$. Lack of Cu atoms in the ico-like cluster occurs at $\alpha > 0.2$; the lack is non monotonically depends on the $\alpha$ value. The minimum value of $\delta N_{\rm Cu}(\alpha)$ occurs at $\alpha = 0.45$ with $\delta N_{\rm Cu}(\alpha)/N_{\rm cl} \approx -0.12$, where  $N_{\rm cl} = 13$ is the number of atoms in the cluster. In the range of interest of copper abundance ($\alpha \in 0.6 \div 0.8$), values of $\delta N_{\rm Cu}(\alpha)/N_{\rm cl} \simeq -0.08$. It is also seen that local extrema of $\delta N_{\rm Cu}(\alpha)$ curve occur at the same $\alpha=0.645, 0.785$ as those on $n_{\rm ico}(\alpha)$ dependence. That means that abundance of ico-like clusters and their chemical composition are correlated.

\section{Conclusions}

To conclude we investigated the local orientational order of the simulated $\rm Cu_{\alpha}Zr_{1-\alpha}$ metallic glass in fine detail by using bond order parameter method and found out the non monotonic behavior of the abundance of ico-like cluster versus $\alpha$ value. Comparison of the BOOP and Voronoi methods reveals that Voronoi method significantly overestimates the number of ico-like particles in the system. We show that abundance of the ico-like clusters and the chemical composition of these clusters are essentially non monotonic versus $\alpha$ and demonstrate local extrema. That qualitatively explains the existence of pinpoint concentrations of high glass-forming ability observing in Cu$\--$Zr alloys. Finally, it has been shown that VT method overestimates drastically the abundance of the ico-like clusters in comparison with the BOOP one.

\section{Acknowledgments}

MD Simulations and local orientational order analysis were supported by the Russian Science Foundation, (grant RNF No. 14-50-00124). Structural analysis was supported by Russian Science Foundation (grant Nr. 14-12-01185). BAK was supported by the A*MIDEX grant (Nr.~ANR-11-IDEX-0001-02) funded by the French Government ``Investissements d'Avenir'' program. Institute of metallurgy thanks UB RAS for access to "Uran" cluster.

\bibliography{klumov}
\end{document}